\documentclass[prb,aps,superscriptaddress,showpacs,twocolumn]{revtex4}
\usepackage{graphicx,amsfonts,bm,amsmath}
\newcommand{\kk}{{\bm k}}
\newcommand{\rl}{\rangle\!\langle}

\newcommand{\ff}{{\cal F}}
\newcommand{\hh}{{\cal G}}
\newcommand{\rr}{{\bf r}}

\DeclareMathOperator{\re}{Re}

\begin{document}

\author{A. Grodecka}
\email{anna.grodecka@uni-paderborn.de}
\affiliation{Computational Nanophotonics Group, Theoretical Physics, 
University Paderborn, 33098 Paderborn, Germany}
\author{P. Machnikowski}
\affiliation{Institute of Physics, Wroc{\l}aw University of Technology,
50-370 Wroc{\l}aw, Poland}
\author{J. F{\"o}rstner}
\affiliation{Computational Nanophotonics Group, Theoretical Physics, 
University Paderborn, 33098 Paderborn, Germany}

\title{Phonon-assisted tunneling between singlet states in~two-electron quantum dot molecules} 

\begin{abstract}
We study phonon-assisted electron tunneling in semiconductor quantum
dot molecules. In particular, singlet-singlet relaxation in a
two-electron doped structure is considered. The influence of Coulomb
interaction is discussed via comparison with a single electron
system. We find that the relaxation rate reaches similar values in the
two cases but the Coulomb interaction shifts the maximum rates towards
larger separations between the dots. The difference in electron-phonon
interaction between deformation potential and piezoelectric coupling
is investigated. We show that the phonon-induced tunneling between
two-electron singlet states is a fast process, taking place on the
time scales of the order of a few tens of picoseconds.  
\end{abstract}

\pacs{73.21.La, 03.65.Yz, 63.20.kd}

\maketitle

\section{Introduction}

Coupled quantum dots (QDs), often referred to as quantum dot molecules
(QDMs), have recently attracted much attention \cite{bayer01,petta05}
due to their potential 
application in various implementations of quantum computation
schemes. Specifically, there have been many proposals for employing
two-electron spin states in QDMs \cite{loss98,barrett06,hanson07}, benefiting
from long decoherence times of the spin \cite{kroutvar04}. For
instance, it was suggested to use singlet and triplet states as logical
qubit states and to perform quantum computation \cite{taylor05} and
Bell-states measurements \cite{Zhang07}. Initialization, control and
read-out of the state of two confined electrons in a QDM have already
been experimentally demonstrated \cite{petta05}. Moreover,
such structures are proposed for coherent optical manipulation of
two-electron states \cite{emary07c,hakan07}.

Semiconductor QDMs are embedded in a solid state
environment, which leads to electron interaction with the phonon
reservoir.
The presence of phonon-mediated coupling between 
energy eigenstates of a QDM leads to new effects in the
physics of these structures, 
as compared to individual QDs \cite{muljarov05,rozbicki08}.
In particular, if the lowest states correspond to electron localization
in two different dots, the relaxation between these states has the
character of phonon-assisted tunneling, consisting in the transfer of
an electron from one dot to the other. Such a process results from an
interplay between the carrier-phonon coupling and tunneling coupling
between the dots, which is a desirable element of many proposals of
QDM-based quantum computing. 
Phonon-assisted tunneling has
been thoroughly studied in the case of QDMs doped with a single
electron \cite{wu05,stavrou05,vorojtsov05,lopez05}. Also
phonon-induced triplet-singlet relaxation (via spin-orbit coupling)
has been analyzed \cite{climente07}. However, to our knowledge,
spin-conserving relaxation between singlet 
states of a two-electron system has not been discussed. 

In this paper, we analyze phonon-assisted tunneling in a quantum dot
molecule consisting of two laterally coupled semiconductor quantum
dots. A system doped with two electrons is considered. 
We study singlet-singlet relaxation, that is, relaxation between
the two lowest states of two electrons in a QDM corresponding to the
singlet spin configuration.
For a specific GaAs QDM system, phonon-assisted relaxation rates are calculated.
As we will show, in the parameter areas where the relaxation is
efficient, it involves charge transfer between the dots. Thus, it
represents a phonon-assisted tunneling process.
We study how the Coulomb interaction in the two-electron system
influences the relaxation of electrons in comparison with the
case of a QDM doped with a single electron. It is demonstrated that
the presence of one electron strongly affects the tunneling of the
other. As a result, the rates of the phonon-assisted electron tunneling
for the two doping cases (with one or two electrons) differ
considerably, which is especially apparent in their dependence on the
distance between the constituent QDs. We consider electrons
interacting with acoustic phonon modes via deformation potential and
piezoelectric couplings and show their distinguished impact on
tunneling in QDMs. It is shown that the piezoelectric mechanism
resulting from the considerable change of charge distribution is of
great importance in the considered system and for some ranges of QDM
parameters it is even the dominant contribution to relaxation. We show
that the phonon-assisted tunneling is strong in coupled quantum dots
and one should be aware of its influence when designing quantum
computation schemes in QDMs.

The paper is organized as follows. In Sec.~\ref{sec:model}, we
introduce the model describing a quantum dot molecule with the Coulomb
interaction and coupling to the phonon
environment. Section~\ref{sec:results} contains the results on
phonon-assisted tunneling rates for the two systems under
consideration. In Sec.~\ref{sec:conclusion}, we conclude the paper
with final remarks. In the Appendix, we summarize the theory of
single-electron phonon-assisted tunneling.

\section{Model}\label{sec:model}

\subsection{Electron states}

\begin{figure}[tb]
\begin{center} 
\unitlength 1mm
{\resizebox{38mm}{!}{\includegraphics{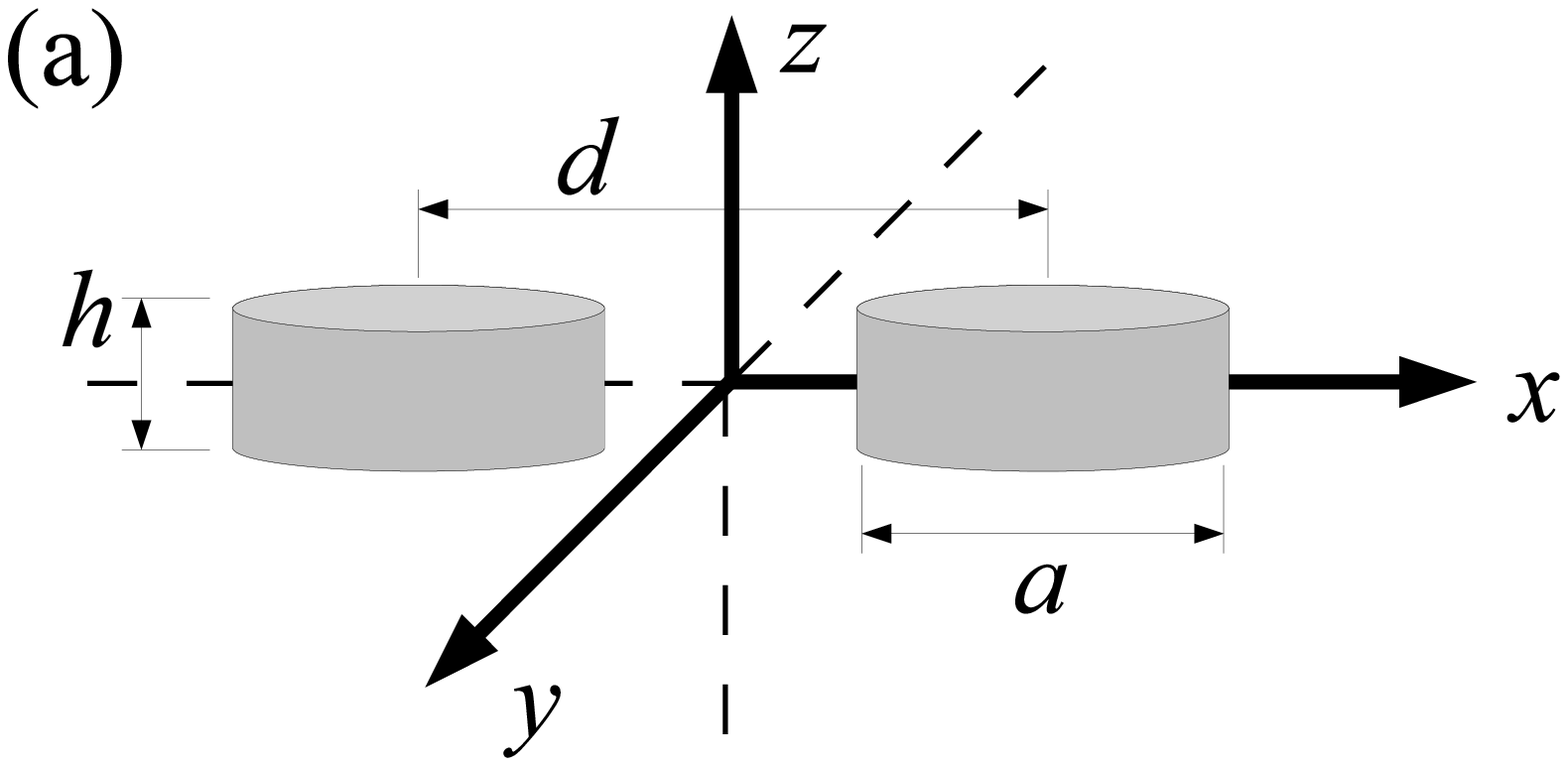}}}
{\resizebox{47mm}{!}{\includegraphics{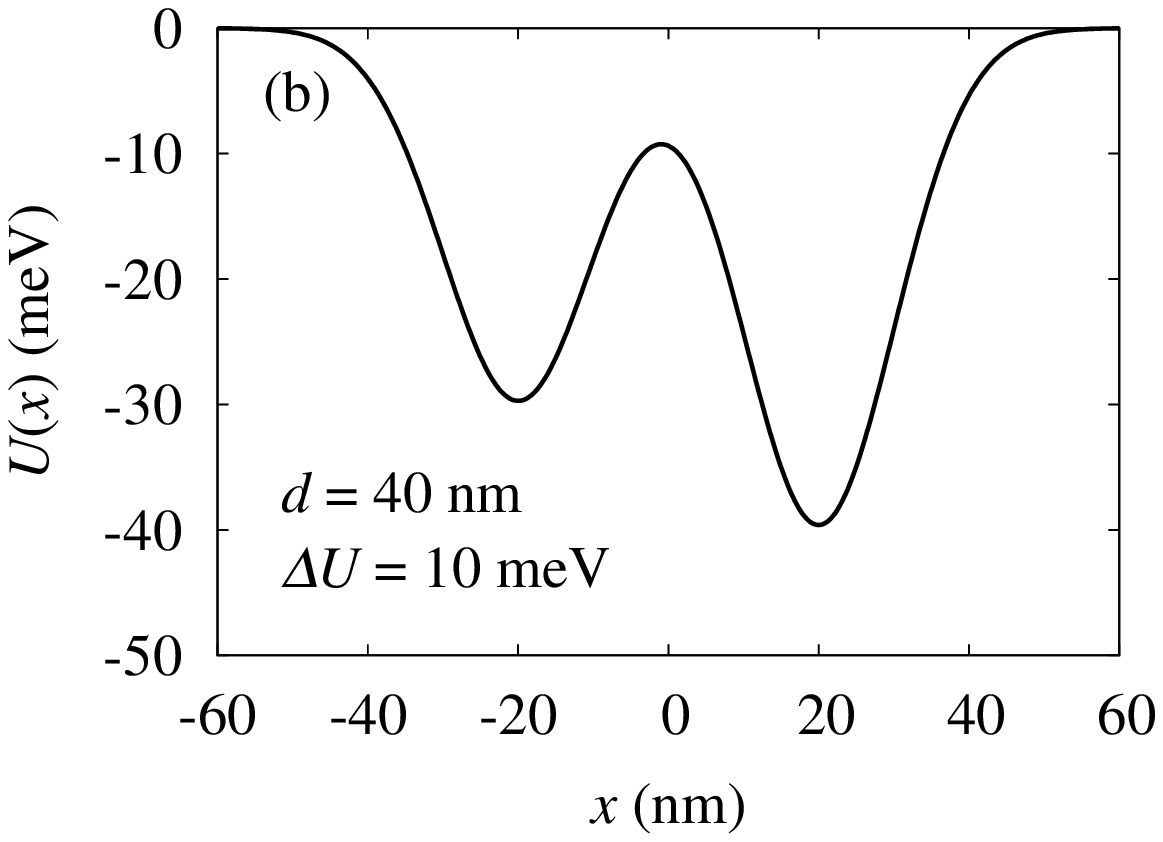}}}
\end{center} 
\caption{\label{fig:scheme} (a) Schematic plot of a laterally coupled double quantum dot. (b) Lateral confinement potential describing the double QD structure.}
\end{figure}

We consider a quantum dot molecule which consists of two laterally (in
$x$ direction) coupled quantum dots
[see~Fig.~\ref{fig:scheme}(a)]. The structure doped with two electrons
is studied. The Hamiltonian of the electron subsystem is given by
\begin{equation}\label{eq:ham}
H_{\rm e} = \frac{\hbar^{2}}{2 m^{*}}\left( \nabla_{\rm a}^{2} +
\nabla_{\rm b}^{2} \right) + U(\rr_{\rm a}) + U(\rr_{\rm b})+
V(\rr_{\rm a},\rr_{\rm b}),
\end{equation}
where $m^{*} = 0.07 m_{0}$ is the effective mass of an electron in
GaAs. $U(\rr_{\rm a/b})$ is the confinement potential for two
electrons referred to as `a' and `b', respectively, and 
\begin{equation*}
V (\rr_{\rm a},\rr_{\rm b}) = \frac{e^{2}}{4 \pi \varepsilon_{0}
\varepsilon_{\rm r}}
\frac{1}{|\rr_{\rm a}-\rr_{\rm b}|}
\end{equation*}
is the Coulomb interaction between the electrons.
Here, $e$ denotes electron charge, $\varepsilon_{0}$ is the vacuum
dielectric constant, and $\varepsilon_{\rm r}$ is the static relative
dielectric constant.

We assume a separable confinement potential
\begin{eqnarray}
U(\rr) & = & U(z) + U(y) + U(x) \\ \nonumber & = & \frac{1}{2} m^{*}
\omega_{z}^{2} \; z^{2} + \frac{1}{2} m^{*} \omega_{y}^{2} \; y^{2} +
U(x),
\end{eqnarray}
where $U(z)$ is the potential in the growth direction and $U(y)$ is
the lateral confinement potential. The potential describing the double
quantum dot structure is $U(x)$ and has two minima, defining the two
QDs. We choose it in the form:
\begin{eqnarray}
U(x) & = & - U_{0} \exp \left[ - \frac{1}{2} \left(
\frac{x - d/2}{a} \right)^{2} \right] \\ \nonumber
&& - (U_{0} + \Delta U) \exp \left[ - \frac{1}{2} \left(
\frac{x + d/2}{a} \right)^{2} \right].
\end{eqnarray}
This model potential has the advantage that it is smooth and allows
one to
independently control the distance between the dots $d$ and the
depths of both potential wells, $U_{0}$ and $U_{0} + \Delta U$. The
difference between the depths of the two constituent dots, $\Delta U$,
is referred to as the \textit{offset}.

The dynamics in the growth and lateral $y$ directions is restricted to
the respective ground states, which are described by Gaussian wave
functions
\begin{eqnarray}
\phi(z) & = & \frac{1}{\sqrt{h \sqrt{\pi}}} \exp \left( -\frac{z^{2}}{2 h^{2}} \right), \\
\varphi(y) & = & \frac{1}{\sqrt{l \sqrt{\pi}}} \exp \left( -\frac{y^{2}}{2 l^{2}} \right).
\end{eqnarray}
Here, $h$ denotes the electron wave function width in the growth direction
$z$, while $l$ is the width in the lateral direction $y$. The restriction
to the ground states is a
reasonable assumption in the considered confinement conditions, since
the energy separation from the next eigenstates has typical values
larger that $10$~meV so that these states do not contribute to the studied dynamics.
The complete wave function of a single electron can be written in a
product form
\begin{equation}\label{eq:psi}
\Phi_{n}(\rr) = \psi_{n}(x) \; \varphi(y) \; \phi(z),
\end{equation}
where $\psi_{n}(x)$ is the $n$th lowest state of electron in a QDM
obtained together with its eigenenergies from the numerical solution
of Schr{\"o}dinger equation. The two lowest single particle
eigenstates are described by the wave functions $\Phi_{0}(\rr)$ and
$\Phi_{1}(\rr)$ with the corresponding energies $\epsilon_{0}$
and~$\epsilon_{1}$.

In order to analyze the relaxation mechanisms for a system doped with
two electrons, we construct two-particle spin-singlet states 
\begin{eqnarray}
|\mathrm{RL} \rangle & = & \frac{1}{\sqrt{2}}\left(
a_{0\uparrow}^{\dag} a_{1\downarrow}^{\dag} + a_{1\uparrow}^{\dag}
a_{0\downarrow}^{\dag} \right) |{\rm vac} \rangle, \\ |\mathrm{RR}
\rangle & = & a_{0\uparrow}^{\dag} a_{0\downarrow}^{\dag} |{\rm vac}
\rangle.
\end{eqnarray}
Here, $ a_{0(1)\uparrow (\downarrow)}^{\dag}$ creates an electron in the
$0$th ($1$st) lowest single-particle state with spin up (spin down), and $|{\rm
vac}\rangle$ denotes an empty quantum dot system. The corresponding
spatially symmetric wave functions are
\begin{eqnarray}
\Psi_{\mathrm{RL}}(\rr_{\rm a}, \rr_{\rm b}) & = & \frac{
\Phi_{0}(\rr_{\rm a}) \Phi_{1}(\rr_{\rm b}) 
+ \Phi_{1}(\rr_{\rm a}) \Phi_{0}(\rr_{\rm b})}{\sqrt{2}}, \\ 
\Psi_{\mathrm{RR}}(\rr_{\rm a}, \rr_{\rm b}) & = & \Phi_{0}(\rr_{\rm
a}) \Phi_{0}(\rr_{\rm b}). 
\end{eqnarray}


For the considered two-electron system, we include the Coulomb
interaction between electrons and solve the secular equation 
in the subspace spanned by the states $|\mathrm{RL}\rangle$ and 
$|\mathrm{RR}\rangle$, with the projected Hamiltonian
\begin{equation*}
\tilde H = 
\left( \begin{array}{cc}
    \epsilon_{0} + \epsilon_{1} + v_{00} & v_{01} \\ v_{01} & 2
    \epsilon_{0} + v_{11} \end{array}\right),
\end{equation*}
where the Coulomb matrix elements are
\begin{eqnarray}\nonumber
v_{00} & = & V_{0} \int d^3 \kk \; \frac{a}{k^{2}}
\left\{ \re \left[ \ff_{00}^{*}(\kk) \ff_{11}(\kk) \right]
+ |\ff_{01}(\kk)|^{2} \right\}, \\ v_{01} & = & \sqrt{2} V_{0} \int
d^3 \kk \; \frac{a}{k^{2}}
\re \left[ \ff_{00}^{*}(\kk) \ff_{01}(\kk) \right], \\
v_{11} & = & V_{0} \int d^3 \kk \; \frac{a}{k^{2}}
|\ff_{00}(\kk)|^{2},
\end{eqnarray}
with
\begin{equation*}
V_{0} = \frac{e^{2}}{8 \pi^{3} \varepsilon_{0} \varepsilon_{\rm r} a}.
\end{equation*}
The single-particle form factors are defined as
\begin{equation}\label{eq:ff}
\ff_{nm}(\kk) = \int d^{3}\rr \; \Phi_{n}^{*}(\rr) e^{i\kk \rr} \Phi_{m}(\rr)
\end{equation}
and for our choice of Gaussian wave functions in the $y$ and $z$
directions are
\begin{eqnarray}
\ff_{nm}(\kk) & = & \exp{\left( -\frac{k_{z}^{2} h^{2}}{4} -\frac{k_{y}^{2} l^{2}}{4} \right)} \\ \nonumber
&& \times \int dx \; \psi_{n}^{*}(x) e^{i k_{x} x} \psi_{m}(x).
\end{eqnarray}

The resulting eigenstates of the interacting system 
are labeled as $|0\rangle$ and $|1\rangle$ and the
corresponding wave functions can be written in the form
\begin{eqnarray*}
\Psi_{0}& = & \cos\frac{\alpha}{2} \Psi_{\mathrm{RL}}
+ \sin\frac{\alpha}{2} \Psi_{\mathrm{RR}}, \\ 
\Psi_{1} & = & -\sin\frac{\alpha}{2} \Psi_{\mathrm{RL}} 
+ \cos\frac{\alpha}{2} \Psi_{\mathrm{RR}},
\end{eqnarray*}
where 
\begin{equation*}
\alpha = \arctan\left( \frac{v_{01}}{\epsilon_{0}-\epsilon_{1}+v_{11}-v_{00}}
\right),
\end{equation*}
and the energies are $E_{0}$ and $E_{1}$,
respectively. The splitting between the two-electron energies is
\begin{equation*}
\Delta E = |E_{1} - E_{0}| = \sqrt{(\epsilon_{0} - \epsilon_{1} +
v_{11} - v_{00})^{2} + 4 v_{01}^{2}}. 
\end{equation*}


\subsection{Carrier-phonon coupling}

In the considered QDM system, carriers not only interact with each
other, but are also coupled to phonons.

The free phonon Hamiltonian is
\begin{equation*}
H_{\rm ph} = \sum_{s,\kk} \hbar \omega_{s,\kk}^{\phantom{\dag}}
b_{s,\kk}^{\dag} b_{s,\kk}^{\phantom{\dag}},
\end{equation*}
where $b_{s,\kk}^{\dag}$ and $b_{s,\kk}^{\phantom{\dag}}$ denote
phonon creation and annihilation operators, respectively. The
corresponding frequencies are $\omega_{s,\kk}$, where $s$ labels
different phonon branches and $\kk$ is the phonon wave vector.

The interaction of the electrons with the phonon reservoir is
described by the Hamiltonian
\begin{equation}\label{Hint}
H_{\rm int} = \sum_{nm,\sigma} a_{n,\sigma}^{\dag}
a_{m,\sigma}^{\phantom{\dag}}
\sum_{s,\kk} F_{s,nm}(\kk) 
\left(b_{s,\kk}^{\phantom{\dag}} + b_{s,-\kk}^{\dag} \right),
\end{equation}
where $F_{s,nn'}(\kk)$ are the single-particle coupling constants [see
Eqs.~(\ref{FDP}) and~(\ref{FPE}) in the Appendix],
which have the symmetry $F_{s,nn'}(\kk) = F_{s,n'n}^{*}(-\kk)$, and
$\sigma$ denotes the spin orientation.

We consider the electron relaxation in the double-QD structure,
which is a real transition on a picosecond time
scale, therefore it can be treated within the Fermi golden rule approach. The
coupling between the two considered singlet states in a two-electron
configuration, resulting from the carrier-phonon interaction
Hamiltonian (\ref{Hint}) is
\begin{equation*}
H_{\rm int} = |0 \rl 1| \sum_{s,\kk} G_{s}(\kk)
\left(b_{s,\kk}^{\phantom{\dag}} + b_{s,-\kk}^{\dag} \right) + \mathrm{H.c.},
\end{equation*}
where $G_{s}(\kk)$ are the two-electron coupling constants (given
below).

The energy difference between the electron states is considerably
smaller than the energy of longitudinal optical phonons (LO), which is
$36$~meV in GaAs, thus they will not contribute to the relaxation
mechanisms. Therefore, we consider interaction only with the relevant
acoustic phonons via the deformation potential and the piezoelectric
coupling.

Using the carrier-phonon coupling constant for the deformation
potential interaction [Eq. (\ref{FDP})], one finds the effective coupling between
the two-electron states
\begin{equation*}
G^{\rm DP}_{{\rm l}}(\kk) = \sqrt{\frac{\hbar k}{2 \rho V c_{\rm l}}}
D_{\rm e} \hh (\kk),
\end{equation*}
where $\rho$ is the crystal density, $V$ is the normalization volume
of the phonon modes, $c_{\rm l}$ is the longitudinal speed of sound,
and $D_{\rm e}$ is the deformation potential constant for
electrons. The geometrical properties of the wave functions are
reflected in the form factor $\hh (\kk)$, which has the form
\begin{eqnarray}
\hh (\kk) & = & \int d^{3}\rr_{\rm a} \int d^{3}\rr_{\rm b} 
\Psi_{0}^{*}(\rr_{\rm a}, \rr_{\rm b}) \\ \nonumber
&& \times \left(e^{i\kk \rr_{\rm a}} + e^{i\kk \rr_{\rm b}} \right)
\Psi_{1}(\rr_{\rm a}, \rr_{\rm b}).
\end{eqnarray}
It can be written by means of the single-particle form factors defined
by Eq.~(\ref{eq:ff}) as
\begin{eqnarray}\nonumber
\hh (\kk) & = & \frac{1}{2} \sin \alpha \left[ \ff_{00}(\kk) - \ff_{11}(\kk) \right]
+ \sqrt{2} \cos \alpha \; \ff_{01}(\kk) \\ & = & \exp{\left(
-\frac{k_{z}^{2} h^{2}}{4} -\frac{k_{y}^{2} l^{2}}{4} \right)} \;
\tilde \hh (k_{x}).
\end{eqnarray}

The two-particle coupling elements for the piezoelectric interaction
are
\begin{equation}\label{eq:piez}
G^{\rm PE}_{s}(\kk) = -i \sqrt{\frac{\hbar}{2 \rho V c_s k}}
\frac{d_{\rm P} e}{\varepsilon_{0}\varepsilon_{\rm r}} M_s (\hat\kk) \hh (\kk),
\end{equation}
where $c_s$ is the speed of sound (longitudinal $c_{\rm l}$ or
transverse $c_{\rm t}$, depending on the phonon branch) and $d_{\rm
P}$ is the piezoelectric constant. The function $M_s (\hat\kk)$ does
not depend on the value of the phonon wave vector, but only on its
orientation. For a zinc-blende structure, it reads
\begin{eqnarray} \nonumber
M_s (\hat\kk) & = & \hat k_{x}
\left[ (\hat e_{s,\kk})_{y}\hat k_{z} + (\hat e_{s,\kk})_{z}\hat k_{y} \right] \\ \nonumber
&& + \hat k_{y} \left[ (\hat e_{s,\kk})_{z}\hat k_{x} + (\hat
e_{s,\kk})_{x}\hat k_{z} \right] \\ && + \hat k_{z} \left[ (\hat
e_{s,\kk})_{x}\hat k_{y} + (\hat e_{s,\kk})_{y}\hat k_{x} \right],
\end{eqnarray}
where $\hat e_{s,\kk}$ is the unit polarization vector for the phonon
wave vector $\kk$ and polarization $s$, and $\hat\kk = \kk/k$.  We
choose the following phonon polarization vectors
\begin{eqnarray}
\hat e_{{\rm l},\kk} & \equiv & \hat\kk = \left( 
\cos \theta, \sin \theta \cos \varphi, \sin \theta \sin \varphi \right),\\ \nonumber
\hat e_{{\rm t1},\kk} & = & \left( 
0, \sin \varphi, - \cos \varphi \right),\\ \nonumber
\hat e_{{\rm t2},\kk} & = & \left( 
-\sin \theta, \cos \theta \cos \varphi, \cos \theta \sin \varphi
\right),
\end{eqnarray}
for which the functions $M_{s}(\hat\kk)$ read
\begin{eqnarray}\label{Ms}
M_{\rm l} (\hat\kk) & = & \frac{3}{2} \sin\theta \sin(2\theta)
\sin(2\varphi), \\ \nonumber M_{\rm t1} (\hat\kk) & = & -
\sin(2\theta) \cos(2\varphi), \\ \nonumber M_{\rm t2} (\hat\kk) & = &
\sin\theta \left(3 \cos^{2} \theta -1 \right) \sin(2\varphi).
\end{eqnarray}


The properties of the phonon environment are represented by phonon
spectral density
\begin{eqnarray}
R (\omega) & = & \frac{1}{\hbar^{2}} |n(\omega)+1| \sum_{s,\kk}
|G_{s}(\kk)|^{2} \\ \nonumber && \times \left[
\delta(\omega-\omega_{s,\kk}) + \delta(\omega+\omega_{s,\kk}) \right],
\end{eqnarray}
where $n(\omega)$ is the Bose distribution function. The deformation
potential contribution is
\begin{eqnarray}
R^{\rm DP}(\omega) & = & R_{0}^{\rm DP} \omega^{3} \; |n(\omega)+1| \\
\nonumber && \times \int_{0}^{2\pi} d\varphi \int_{0}^{\pi} \sin
\theta \; d \theta \; |\tilde\hh (\omega \cos\theta/c_{\rm l})|^{2}\\
\nonumber && \times \exp{\left[ -\frac{\omega^{2}}{2c_{\rm l}^{2}}
\sin^{2}\theta \left( l^{2} \cos^{2} \varphi + h^{2} \sin^{2} \varphi \right) \right]},
\end{eqnarray}
where
\begin{equation*}
R_{0}^{\rm DP} = \frac{D_{\rm e}^{2}}{16 \pi^{3} \hbar \rho c_{\rm
l}^{5}}.
\end{equation*}
The piezoelectric term is
\begin{eqnarray*}
R_s^{\rm PE}(\omega) & = & R_{0,s}^{\rm PE} \; \omega \; |n(\omega)+1|
\int_{0}^{2\pi} d\varphi \int_{0}^{\pi} \sin \theta \; d \theta \\ \nonumber
&& \times |M_s(\hat\kk(\varphi,\theta))|^{2} |\tilde\hh (\omega
\cos\theta/c_s)|^{2} \\ \nonumber && \times \exp{\left[
-\frac{\omega^{2}}{2c_s^{2}}
\sin^{2}\theta \left( l^{2} \cos^{2} \varphi + h^{2} \sin^{2} \varphi \right) \right]},
\end{eqnarray*}
where
\begin{equation*}
R_{0,s}^{\rm PE} = \frac{d_{\rm P}^{2} e^{2}} {16 \pi^{3} \hbar \rho
c_s^{3} \varepsilon_{0}^{2} \varepsilon_{\rm r}^{2}}.
\end{equation*}
Note that the coupling constants for deformation potential and
piezoelectric channels have different parity (as functions of $\kk$)
so that these two transition channels do not interfere.

In order to study phonon-assisted relaxation, we employ the Fermi
golden rule and obtain the relaxation rate
\begin{equation}\label{eq:rate}
w = 2\pi R\left(\frac{\Delta E}{\hbar}\right),
\end{equation}
which is proportional to the phonon spectral density at the frequency
corresponding to the splitting energy $\Delta E$.

\begin{table}[tb]
\begin{tabular}{lll}
\hline\hline
Deformation potential for electrons \;\;\;\;\;\; & $D_{e}$ & $-8.0$ eV \\ 
Density & $\rho$ & 5360 kg/m$^{3}$ \\ 
Longitudinal sound speed & $c_{\rm l}$ & 5150 m/s \\ 
Transverse sound speed & $c_{\rm t}$ & 2800 m/s \\ 
Static dielectric constant & $\varepsilon_{\rm r}$ & 13.2 \\ 
Piezoelectric constant & $d_{\rm P}$ \;\;\; & 0.16 C/m$^{2}$ \\
Confinement depth & $U_{0}$ & 30 meV \\
Wave-function width in: \\
\;\;\;\;\;\; $z$-direction \;\;\;\;\;\; & $h$ & 4.0~nm \\
\;\;\;\;\;\; $y$-direction \;\;\;\;\;\; & $l$ & 10.0~nm \\
\;\;\;\;\;\; $x$-direction \;\;\;\;\;\; & $a$ & 10.0~nm \\
\hline\hline
\end{tabular}
\caption{\label{tab:param}The GaAs material parameters and QDM system parameters.}
\end{table}

The material parameters (corresponding to GaAs quantum dots) and
parameters of the QDM system are given
in~Table~\ref{tab:param}. Moreover, details concerning relaxation
in a single electron QDM system, which will be used for
comparison, are presented in the Appendix.


\section{Results: tunneling rates}\label{sec:results}

In this section, the results for phonon-assisted transitions in a double
quantum dot are presented. We consider a QDM doped with two electrons
and the singlet-singlet relaxation channel. In order to investigate how the
Coulomb interaction influences the relaxation processes, the results are compared
with those for a single electron case, calculated in a way similar to
Refs. \onlinecite{wu05,stavrou05,vorojtsov05,lopez05} (see Appendix). The
quantitative results are obtained at temperature $T=0$ K for GaAs
quantum dots with the sizes $h = 4$~nm and $l = a = 10$~nm in growth
and lateral directions, respectively.

The probability of phonon-assisted electron transitions
[Eq.~(\ref{eq:rate})] is proportional to the spectral density of the
phonon reservoir at the frequency corresponding to the energy
splitting $\Delta E$. Therefore, the transition rate will be high when
this energy lies in the frequency range of maximal values of the
phonon spectral density. In order to see which parameter range is
favorable for relaxation, we first study the energy splittings and
phonon spectral densities for the two considered doping cases.

\begin{figure}[tb]
\begin{center} 
\unitlength 1mm
{\resizebox{80mm}{!}{\includegraphics{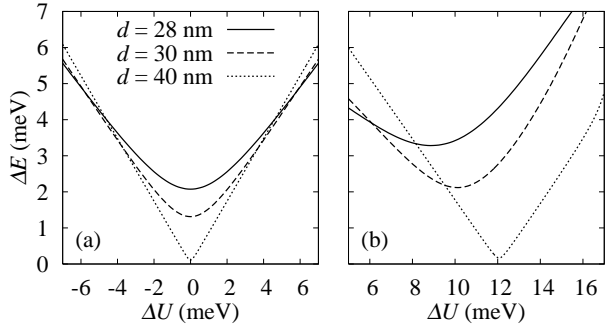}}}
\end{center} 
\caption{\label{fig:split} (a) Splitting energy as a function of the QDs offset for a single electron in a QDM for different distances $d$ between QDs. (b) As in (a) but for two electrons in a QDM.}
\end{figure}

In Fig.~\ref{fig:split}(a), we plotted the energy splitting for a
single electron in a QDM as a function of the confinement depth offset
$\Delta U$ for $U_{0} = 30$~meV and a few values of the distance $d$ between the
QDs. The minimum value occurs always 
when the QDs are the same and, in general, is smaller for larger
distances between dots, where they do not influence each other. In the
case of two electrons in a QDM [Fig.~\ref{fig:split}(b)], the
splitting energies have a slightly more complicated behavior. Now, the
minimum value is shifted due to interplay between the on-site
(single-particle) potential and the Coulomb interaction, which also
depends on the distance between QDs. The splitting energies are
larger, since they describe two-particle states affected by the Coulomb
coupling. While for one electron, the energies are symmetric with
respect to the resonance point (minimum splitting), 
in the two-electron case this symmetry is lost, except for very large separations.

\begin{figure}[tb]
\begin{center} 
\unitlength 1mm
{\resizebox{80mm}{!}{\includegraphics{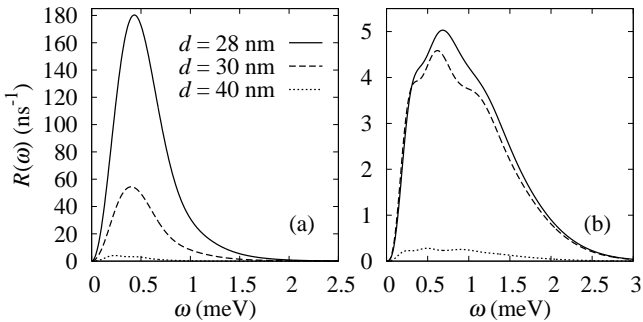}}}
\end{center} 
\caption{\label{fig:spectral} (a) Total spectral density of the phonon
reservoir for one electron in a QDM for $\Delta U = 0$~meV and different distances $d$ 
between the QDs. (b) As in (a) but
for two electrons and $\Delta U = 11$~meV.} 
\end{figure}

Since the wave functions obviously depend on the confinement
offset $\Delta U$, so do the coupling constants  and, in consequence,
the phonon spectral densities. In order to gain some information on
their general shape, we study the spectral densities for the 
values of the offset which correspond to the minimal values of energy
splitting. In Fig.~\ref{fig:spectral}(a), the phonon spectral density for a
single electron [see Eqs.~(\ref{rrdp}) and~(\ref{rrpe}) in the Appendix] 
is plotted for the offset $\Delta U = 0$~meV. In general, 
the values of phonon spectral densities depend on the overlap between
the wave functions and thus are large for small distances $d$ between
the QDs. The function has its maximum for $\omega \approx 0.4$~meV and
a cut-off at $\omega \approx 2.5$~meV. One can expect high
rates for energy splittings from $0.1$ to $1$~meV, especially for
small distances $d$. From Fig.~\ref{fig:split}(a) it is clear that for
closely spaced QDs, the energy splitting is larger than $2$~meV and
lies almost beyond the cut-off of the phonon density, which will
result in lower transition rates. For larger distances the splitting is
smaller, but also the amplitude of the spectral density is
smaller. The interplay of phonon density and splitting energies will
be reflected in nontrivial dependence of relaxation rates on the
distance between the QDs.

For a two-electron QDM, the phonon spectral density has, in general,
smaller values [Fig.~\ref{fig:spectral}(b)], since the overlap between
corresponding two-electron wave functions is smaller. In this case,
the cut-off energy ($\omega \approx 3$~meV) as well as the energy
splitting is larger. One can see that phonon-assisted transitions in
both systems will be large for energy splittings smaller than $3$~meV
and will strongly depend on the distance $d$.

\begin{figure}[tb]
\begin{center} 
\unitlength 1mm
{\resizebox{80mm}{!}{\includegraphics{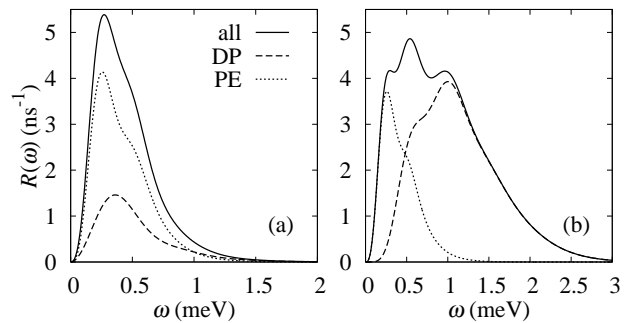}}}
\end{center} 
\caption{\label{fig:wklady} (a) Total phonon spectral density with the
contributions resulting from deformation potential and piezoelectric
couplings for one electron in a QDM for $\Delta U = 0$~meV and $d = 38$~nm. 
(b) As in (a) but for two electrons and $\Delta U = 11$~meV.} 
\end{figure}

The electron-phonon interaction via both deformation potential as well
as piezoelectric coupling is considered next. In order to see which
interaction has a stronger influence, in Figs.~\ref{fig:wklady}(a)
and~\ref{fig:wklady}(b) we present the total spectral density of the
phonon environment together with the two contributions for a fixed
distance $d = 38$~nm. It is clear that piezoelectric coupling in
double quantum dot structures is of great importance in contrast to
optical processes in
single QD structures, where this interaction can in many cases be
neglected \cite{krummheuer02,forstner03,grodecka07}. This results from
the fact that electron relaxation induces a large change of charge
redistribution, especially when it involves tunneling to the other
dot. Since in a single electron as well in a two-electron
system the two phonon contributions may cover different frequency
sectors, they will also play a role in the transition rates in
distinct parameter areas.

\begin{figure}[tb]
\begin{center} 
\unitlength 1mm
{\resizebox{85mm}{!}{\includegraphics{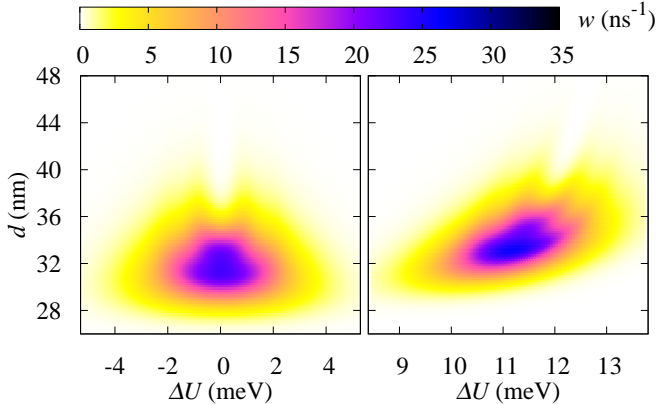}}}
\end{center} 
\caption{\label{fig:mapa-dp} (a) Electron relaxation rate assisted by
phonons via deformation potential coupling for a single electron in a
QDM as a function of QDs offset and distance $d$. (b) As in (a) but
for two electrons.} 
\end{figure}

We start the discussion of phonon-assisted relaxation from the
deformation potential contribution. For a one-electron QDM
[Fig.~\ref{fig:mapa-dp}(a)], the rates are symmetric with respect to
the offset of the quantum dots $\Delta U$. This results from the
symmetric behavior of the splitting energies $\Delta E$. When the QDs
are close to each other, $d\lesssim 28$~nm, the rate is low,
since the energy difference between the two lowest electron states is
much larger than the acoustic phonon energies. Thus one-phonon
transitions are impossible. For distances $d$ from $28$~to~$38$~nm,
the relaxation rate is high. It corresponds to the situation when the
energy splittings are comparable to the phonon energies. The
transition rate reaches its maximum value of~$30$~ns$^{-1}$ for~$d
\approx 32$~nm. Here, the relaxation conditions are most favorable,
since the distance between the QDs is large enough for the
splitting energy to coincide with the maximum value of the
phonon spectral density. For large distances, $d\gtrsim 38$~nm, the
rate vanishes in spite of small splitting energies, since the
overlap between the electron wave functions tends to zero and, in
consequence, the spectral density vanishes. The transition rates are
also small for large offsets, $|\Delta U| \gtrsim 3$~meV, since it
leads to large energy gap between the levels.

For a two-electron QDM [Fig.~\ref{fig:mapa-dp}(b)], the maximum of the
relaxation rate shifts with growing distance towards larger confinement
offsets,
which was already visible in the splitting energies. Larger distances
$d$ between the QDs are needed for efficient relaxation, which is an
evidence of the Coulomb interaction between two electrons, 
leading to an increase of the splitting
energies. In general, the maximum magnitude of the relaxation rates is
comparable with that for a single electron but the parameter range
in which their values are maximal is shifted due to the
electron-electron interaction.

\begin{figure}[tb]
\begin{center} 
\unitlength 1mm
{\resizebox{85mm}{!}{\includegraphics{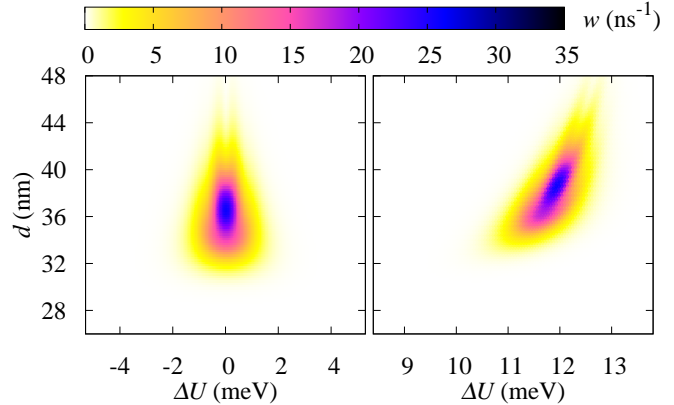}}}
\end{center} 
\caption{\label{fig:mapa-pe} (a) Electron relaxation rate assisted by
phonons via piezoelectric coupling for a single electron in a QDM as a
function of QD confinement offset $\Delta U$ and distance $d$. (b) As in (a) but
for two electrons.} 
\end{figure}

In the case of piezoelectric coupling [Figs.~\ref{fig:mapa-pe}(a)
and~\ref{fig:mapa-pe}(b)], the relaxation rate has relatively large
values in a smaller range of QD offsets. This is a result of a
different form of corresponding spectral density, which, in general,
is narrower than for the deformation potential. Therefore, smaller
splitting energies are more favorable. For the same reason, it is
shifted towards larger distances $d$. The relaxation rate reaches the
values of $30$~ns$^{-1}$, which is as large as that for deformation
potential. This maximum appears at the distance  
$d \approx 36$~nm for a single electron and $d
\approx 39$~nm for a double electron QDM.

\begin{figure}[tb]
\begin{center} 
\unitlength 1mm
{\resizebox{85mm}{!}{\includegraphics{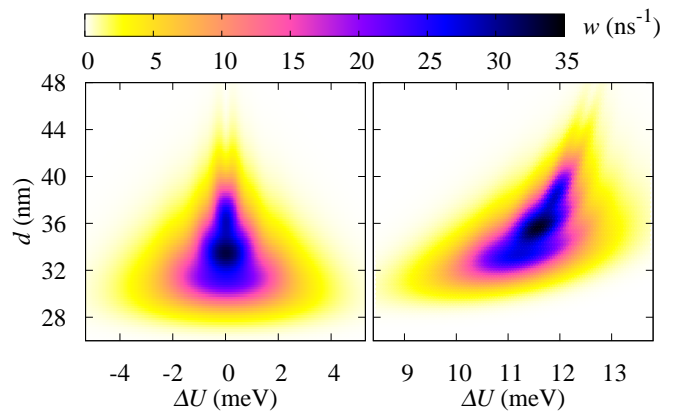}}}
\end{center} 
\caption{\label{fig:mapa-all} (a) Relaxation rate for all phonon modes
and one electron in a QDM. (b) As in (a) but for two electrons.} 
\end{figure}

The total phonon-induced relaxation rate, including both deformation
potential as well as piezoelectric contributions, is
shown in Figs.~\ref{fig:mapa-all}(a)~and~\ref{fig:mapa-all}(b) as a
function of QD offset $\Delta U$ and separation $d$. For a single
electron system, the rate is high for offsets
between $\Delta U = -3$~and~$\Delta U = 3$~meV and for distances from
$d = 28$ to $d = 42$~nm, and reaches its maximal value of
$35$~ns$^{-1}$ for identical QDs separated by the distance of $d
\approx 34$~nm. In 
case of the QDM doped with two electrons, the relaxation mechanism is
strong for offsets between $\Delta U \approx 9$ and $\Delta U \approx
13$~meV and distances from $d \approx 30$ to $d \approx 44$~nm. Its
maximum value also reaches $35$~ns$^{-1}$ for $\Delta U \approx
11.5$~meV and $d \approx 36$~nm.


\begin{figure}[tb]
\begin{center} 
\unitlength 1mm
{\resizebox{85mm}{!}{\includegraphics{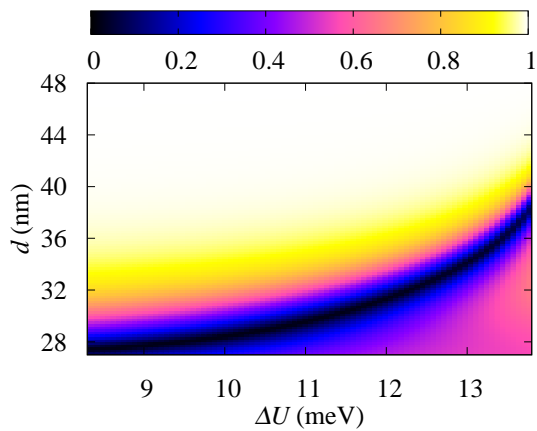}}}
\end{center} 
\caption{\label{fig:alfa} The difference between the average number of electrons 
in the left quantum dot in the two-electron singlet states.} 
\end{figure}

To understand the nature of the relaxation process in the two-electron
case, in Fig.~\ref{fig:alfa}, we plot the difference between the average number 
of electrons in the left quantum dot in the two-electron singlet states. 
One can see that in the area of efficient relaxation
(cf. Fig.~\ref{fig:mapa-all}), the average electron number changes in
most cases almost by one. 
This shows that the relaxation in the
two-electron case is associated with a considerable charge transfer
and, therefore, can be interpreted as a phonon-assisted tunneling process.

For a single electron, the energy eigenstates follow a universal
model of level anticrossing, with the energy splitting 
$\Delta E=\sqrt{(\Delta U)^{2}+4t^{2}}$, where $t$ is the ``tunneling
matrix element'', corresponding to half of the minimum energy splitting
in Fig.~\ref{fig:split}a. This element affects the phonon-assisted
tunneling rate in a twofold way. First, it determines the splitting of
the energy levels and its position with respect to the area of large
phonon spectral density. Second, it affects the degree of mixing of
the wave functions, thus directly changing the spectral density. It
should be noted, however, that the relaxation rate cannot be fully
characterized by this single parameter, since the phonon spectral
density depends on the actual geometry of the system, and therefore
the spatial separation between the dots is itself of direct
importance.
In the two-electron system, the situation is even more complicated,
since the energies and wave functions are affected by the interplay
between the single-particle ``tunnel coupling'' and the Coulomb
interaction between the electrons. This is manifested in the increased
resonance width and loss of symmetry in Fig.~\ref{fig:split}b. As a
result, the relaxation rates are also asymmetric with respect to
$\Delta U$ [see Figs.~\ref{fig:mapa-all}(a) and
\ref{fig:mapa-all}(b)].

\begin{figure}[t]
\begin{center} 
\unitlength 1mm
{\resizebox{80mm}{!}{\includegraphics{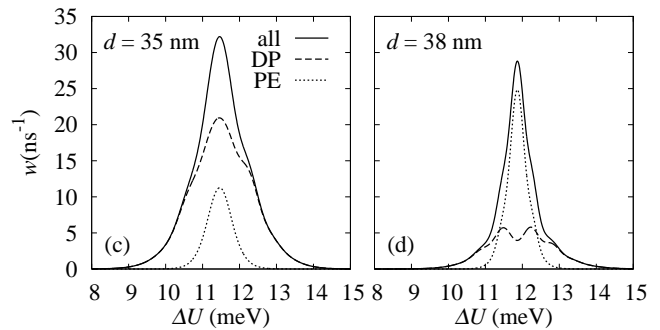}}}
\end{center} 
\caption{\label{fig:przek} (a) Total phonon-assisted relaxation rate with two contributions 
in a two-electron QDM for $d = 35$~nm. (b) As in (a) but for $d = 38$~nm.}
\end{figure}

To have a better insight into particular phonon contributions, in
Figs.~\ref{fig:przek}(a)~and~\ref{fig:przek}(b) we
present the total relaxation rate along with the contributions from
both the coupling mechanisms for a two-electron QDM
for $d = 35$~nm and $d = 38$~nm, respectively. For relaxation rates of
comparable values, the dominant phonon coupling can be different. For
instance, for
$d = 35$~nm, the deformation potential coupling is crucial and the
piezoelectric effect is a few times smaller, while for $d = 38$~nm the
situation is reverse. 


\section{Conclusion}\label{sec:conclusion}

In the present paper, we have studied phonon-assisted relaxation and
tunneling in a 
quantum dot molecule. Structures doped with two electrons have been
considered and compared with the case of a single electron. By
comparison of these two systems, it was shown that the Coulomb
interaction influences the tunneling rates and leads to energy
renormalization and shift of the range of efficient relaxation. We
studied in 
detail carrier-phonon interactions via both deformation potential and
piezoelectric coupling and showed the difference in their behavior and
impact on relaxation. We have shown that the
relaxation in the two-electron case is accompanied by a charge
transfer between the dots and, therefore, can be regarded as a
phonon-assisted tunneling process.

It should be noted that the values of phonon-assisted tunneling rates
in a QDM system are comparable with relaxation times in a single QD
\cite{zibik04}. Moreover, in comparison with the spin coherence times
being up to milliseconds \cite{kroutvar04}, the phonon-assisted
relaxation times are up to several orders of magnitude faster. This
shows that while designing the quantum computer implementations on
electron states in double quantum dots, one has to take into account
the coupling of the carriers to the phonon degrees of
freedom. Finally, it should be noted that the calculations were
performed for zero temperature, which gives a lower bound for
tunneling rates.

\begin{acknowledgments}
A. G. and J. F. acknowledge support from the Emmy Noether Program 
of the Deutsche Forschungsgemeinschaft (Grant No. FO 637/1-1).
P. M. acknowledges support form the Polish MNiSW (Grant No. N202 1336 33). 
P. M. thanks R. Buczko for inspiring discussions.
\end{acknowledgments}

\appendix*

\section{Single electron tunneling}

In this Appendix, the formalism for the tunneling in a single electron
QDM system is presented. In this case, the electron Hamiltonian
[Eq.~(\ref{eq:ham})] is reduced to a simpler form,
\begin{equation*}
H_{\rm e} = \frac{\hbar^{2}}{2 m^{*}} \nabla^{2} + U(\rr),
\end{equation*}
and the electrons are described by a single particle wave functions
given by Eq.~(\ref{eq:psi}). We label the two lowest single electron
states as $|\tilde 0 \rangle $ and $|\tilde 1 \rangle $.

The relevant part of the carrier-phonon interaction Hamiltonian 
describing electron transitions between the constituent QDs is
\begin{equation*}
H_{\rm int} = |\tilde 0 \rl \tilde 1| \sum_{s,\kk} F_{s,01}(\kk)
\left(b_{s,\kk}^{\phantom{\dag}} + b_{s,-\kk}^{\dag} \right) + \mathrm{H.c.},
\end{equation*}
where the single-particle coupling constant for the deformation
potential is
\begin{equation}\label{FDP}
F^{\rm DP}_{{\rm l},01}(\kk) = \sqrt{\frac{\hbar k}{2 \rho V c_{\rm
l}}} D_{\rm e} \ff_{01} (\kk),
\end{equation}
with the form factors given by Eq.~(\ref{eq:ff}). 
The coupling element for piezoelectric interactions reads
\begin{equation}\label{FPE}
F^{\rm PE}_{s,01}(\kk) = -i \sqrt{\frac{\hbar}{2 \rho V c_s k}}
\frac{d_{\rm P} e}{\varepsilon_{0}\varepsilon_{\rm r}} M_s (\hat\kk) \ff_{01} (\kk),
\end{equation}
where the functions $M_s (\hat\kk)$ are given by Eq.~(\ref{Ms}).

The corresponding phonon spectral densities for a single electron in a QDM are
\begin{eqnarray}\label{rrdp}
R_{\rm 1e}^{\rm DP}(\omega) & = & R_{0}^{\rm DP} \omega^{3} \; |n(\omega)+1| \\
\nonumber && \times \int_{0}^{2\pi} d\varphi \int_{0}^{\pi} \sin
\theta \; d \theta \; |\tilde\ff_{01} (\omega \cos\theta/c_{\rm l})|^{2}\\
\nonumber && \times \exp{\left[ -\frac{\omega^{2}}{2c_{\rm l}^{2}}
\sin^{2}\theta \left( l^{2} \cos^{2} \varphi + h^{2} \sin^{2} \varphi \right) \right]}
\end{eqnarray}
and
\begin{eqnarray}\label{rrpe}
R_{\rm{1e},s}^{\rm PE}(\omega) & = & R_{0,s}^{\rm PE} \; \omega \; |n(\omega)+1|
\int_{0}^{2\pi} d\varphi \int_{0}^{\pi} \sin \theta \; d \theta \\ \nonumber
&& \times |M_s(\hat\kk(\varphi,\theta))|^{2} |\tilde\ff_{01} (\omega
\cos\theta/c_s)|^{2} \\ \nonumber && \times \exp{\left[
-\frac{\omega^{2}}{2c_s^{2}}
\sin^{2}\theta \left( l^{2} \cos^{2} \varphi + h^{2} \sin^{2} \varphi \right) \right]},
\end{eqnarray}
where
\begin{equation*}
\tilde \ff_{01} (k_{x}) = \int dx \; \psi_{0}^{*}(x) e^{i k_{x} x} \psi_{1}(x).
\end{equation*}

The Fermi golden rule relaxation rate is then calculated from
Eq. (\ref{eq:rate}), using the total spectral density including both
relaxation channels.

\bibliographystyle{my-prsty}
\bibliography{abbr,quantum}

\end{document}